\documentclass[prd,amsmath,amssymb,showpacs,superscriptaddress,nofootinbib,twocolumn]{revtex4-1}
\usepackage{graphicx}
\usepackage{epsfig,graphics,subfigure,psfrag,amsmath,amssymb}
\usepackage{lineno}
\usepackage{dcolumn}
\usepackage{bm}
\usepackage{overpic}
\usepackage{xspace}
\usepackage{float}

\usepackage{color}
\usepackage{multirow}
\usepackage{colortbl}
\usepackage{epstopdf}
\usepackage[colorlinks,linkcolor=blue,anchorcolor=blue,citecolor=blue]{hyperref}
\usepackage{dcolumn}

\newcolumntype{C}{D{\pm}{\pm}{3, 5}}

\newcommand{\BESIII}{BES\uppercase\expandafter{\romannumeral3}\xspace}

\begin{document}

\title{\boldmath Precision measurements of the $e^+e^- \to
  K_{S}^{0}K^{\pm}\pi^{\mp}$ Born cross sections at center-of-mass
  energies between 3.8 and 4.6 GeV}
\author{
\small
M.~Ablikim$^{1}$, M.~N.~Achasov$^{9,d}$, S. ~Ahmed$^{14}$, M.~Albrecht$^{4}$, M.~Alekseev$^{55A,55C}$, A.~Amoroso$^{55A,55C}$, F.~F.~An$^{1}$, Q.~An$^{52,42}$, Y.~Bai$^{41}$, O.~Bakina$^{26}$
% ...
\\
\vspace{0.2cm}
(BESIII Collaboration)\\
\vspace{0.2cm} {\it
$^{1}$ Institute of High Energy Physics, Beijing 100049, People's Republic of China\\
% ...
$^{4}$ Bochum Ruhr-University, D-44780 Bochum, Germany\\
% ...
$^{9}$ G.I. Budker Institute of Nuclear Physics SB RAS (BINP), Novosibirsk 630090, Russia\\
% ...
$^{14}$ Helmholtz Institute Mainz, Johann-Joachim-Becher-Weg 45, D-55099 Mainz, Germany\\
% ...
$^{26}$ Joint Institute for Nuclear Research, 141980 Dubna, Moscow region, Russia\\
% ...
$^{41}$ Southeast University, Nanjing 211100, People's Republic of China\\
$^{42}$ State Key Laboratory of Particle Detection and Electronics, Beijing 100049, Hefei 230026, People's Republic of China\\
% ...
$^{52}$ University of Science and Technology of China, Hefei 230026, People's Republic of China\\
% ...
$^{55}$ (A)University of Turin, I-10125, Turin, Italy; (B)University of Eastern Piedmont, I-15121, Alessandria, Italy; (C)INFN, I-10125, Turin, Italy\\
% ...
$^{d}$ Also at the Novosibirsk State University, Novosibirsk, 630090, Russia\\
% ...
}
}


\begin{abstract}
  Using data samples collected by the BESIII detector operating at the
  BEPCII storage ring, we measure the $e^+e^- \to
  K_{S}^{0}K^{\pm}\pi^{\mp}$ Born cross sections at center-of-mass
  energies between 3.8 and 4.6\,GeV, corresponding to a luminosity of
  about 5.0~fb$^{-1}$.  The results are compatible with the BABAR
  measurements, but with the precision significantly improved.  A
  simple $1/s^n$ dependence for the continuum process can describe the
  measured cross sections, but a better fit is obtained by an
  additional resonance near 4.2\,GeV,
  which could be an excited charmonium or a charmonium-like state.
\end{abstract}

\maketitle

\section{Introduction}
The charmonium-like state $Y(4260)$ was first observed in the initial
state radiation (ISR) process, $e^+e^- \to \gamma_{\rm{ISR}}\pi^{+}\pi^{-}
J/\psi$, by BABAR\,\cite{Y(4260)_BaBar}, and later confirmed by the
CLEO\,\cite{Y(4260)_CLEO} and Belle\,\cite{Y(4260)_Belle} experiments.
In 2016, a resonant structure, the $Y(4220)$, was observed in the
process $e^+e^- \to \pi^+\pi^- h_{c}$ by the BESIII
collaboration~\cite{pipihc}.  At the same time, BESIII reported a precise
measurement of the $e^{+}e^{-}\to \pi^{+}\pi^{-}J/\psi$ cross sections
in the center-of-mass (c.m.) energy region from 3.77 to 4.60
GeV~\cite{BESIII_pipiJpsi}, where it found the $Y(4260)$ to have a
mass of $(4222.0\pm3.1\pm1.4)$ MeV$/c^{2}$ and a width of
$(44.1\pm4.3\pm2.0)$ MeV, in good agreement with the $Y(4220)$
observed in $e^{+}e^{-}\to\pi^{+}\pi^{-}h_{c}$~\cite{pipihc}. Given
the similar masses and widths, they may be the same particle,
denoted thereafter as $Y(4220/4260)$.
Since $Y(4220/4260)$ is produced in $e^+e^-$ annihilation, its quantum
numbers must be $J^{PC}=1^{--}$. However, $Y(4220/4260)$ seems to have
rather different properties compared with the known charmonium states
with $J^{PC}=1^{--}$ in the same mass region, such as $\psi(4040)$,
$\psi(4160)$ and $\psi(4415)$\,\cite{ccbar_1, ccbar_2, ccbar_3}.
Although above $D\bar{D}$ production threshold, the $Y(4220/4260)$ has
strong coupling to the $\pi^{+}\pi^{-}J/\psi$ final state, instead of
the $D^{(*)}\bar{D}^{(*)}$ final state\,\cite{PiPiJpsi}.  Such a
strong coupling to a hidden-charm final state suggests that the
$Y(4220/4260)$ is a non-conventional $c\bar{c}$ meson.  Various
scenarios have been proposed, which interpret the $Y(4220/4260)$ as a
tetraquark state, hybrid state, molecular state, or dynamical
effect\,\cite{Y(4260)_1, Y(4260)_2, Y(4260)_3, Y(4260)_4, Y(4260)_5},
but all need to be tested with experimental data.  Most previous
studies of the $Y(4220/4260)$ are based on hadronic transitions.  The
CLEO experiment investigated 16 charmonium and light hadron decay
modes based on 13.2~pb$^{-1}$ of $e^+e^-$ data collected at
c.m. energy of $\sqrt{s}=4.260$\,GeV, but only a few decay modes had
significance greater than 3$\sigma$\,\cite{CLEO_2}. The BABAR
collaboration has measured the cross section of $e^+e^- \to
K_{S}^{0}K^{\pm}\pi^{\mp}$\,\cite{BaBar} with the ISR process and
found an excess around $\sqrt{s}=$ 4.2\,GeV, which is very close to
the $\psi(4160)$ and $Y(4220/4260)$. Analyzing this process with a
larger data sample provides higher precision and more information
on $Y(4220/4260)$ decays to light hadrons.

In this paper, we report measurements of the
$e^+e^- \to K_{S}^{0}K^{+}\pi^{-}$, $K_{S}^{0} \to\pi^+ \pi^-$  Born cross section
at c.m. energies from 3.8 to 4.6~GeV. The charge conjugate
decays to $K_{S}^{0}K^{-}\pi^{+}$ are included in this analysis. The
corresponding c.m.  energies\,\cite{ECMS} and the integrated
luminosities\,\cite{Lumi} of all the data samples used in this paper
are summarized in Table~\ref{KsKpi_result}.

\begin{table*}[htbp]
  \begin{center}
    \caption{The measured $e^+e^- \to K_{S}^{0}K^{+}\pi^{-}$ Born
      cross sections. Shown in the table are the integrated
      luminosities $\mathcal{L}$, the numbers of events in the signal
      region $N^{\rm{obs}}$, the numbers of estimated background
      events $N^{\rm{bkg}}$, the signal yields
      $N^{\rm{sig}}=N^{\rm{obs}}-N^{\rm{bkg}}$, the detection
      efficiencies $\epsilon$, the ISR correction factors
      $(1+\delta^{\rm{ISR}})$, the vacuum polarization correction
      factors $\frac{1}{|1-\Pi|^{2}}$ and the measured Born cross
      sections $\sigma_{B}$.  The first uncertainty on the cross
      section is statistical and the second systematic.}

  \label{KsKpi_result}
    \begin{tabular}{ccccccccc}
      \hline \hline
      $\sqrt{s}$~(GeV) &$\mathcal{L}$~(pb$^{-1}$) &$N^{\rm{obs}}$ & $N^{\rm{bkg}}$ & $N^{\rm{sig}}$ &$\varepsilon$~(\%)
      & $(1+\delta^{\rm{ISR}})$ & $\frac{1}{|1-\Pi|^{2}}$ & $\sigma_{B}$(pb)\\
	\hline
	3.808	& 50.1 & 151 & $0.0 $ & 151.0 & 26.4 & 0.901& 1.054  &17.38$\pm 1.41$ $\pm$ 0.77\\
	3.896	& 52.6 & 92  & $1.0 $ & 91.0  & 28.1 & 0.847& 1.047  &10.05$\pm 1.07$ $\pm$ 0.44\\
	4.008	& 480.5& 795 & $11.8$ & 783.2 & 28.8 & 0.844& 1.043  &9.29 $\pm 0.34$ $\pm$ 0.41\\
	4.086	& 52.4 & 78  & $3.0 $ & 75.0  & 27.1 & 0.843& 1.052  &8.62 $\pm 1.04$ $\pm$ 0.38\\
	4.189	& 43.1 & 70  & $1.0 $ & 69.0  & 27.8 & 0.840& 1.056  &9.39 $\pm 1.15$ $\pm$ 0.41\\
	4.208	& 54.3 & 71  & $1.0 $ & 70.0  & 27.1 & 0.840& 1.057  &7.75 $\pm 0.94$ $\pm$ 0.34\\
	4.217	& 54.2 & 80  & $2.0 $ & 78.0  & 27.8 & 0.840& 1.057  &8.43 $\pm 0.98$ $\pm$ 0.37\\
	4.226	&1041.6& 1343& $25.3$ & 1317.7& 26.9 & 0.840& 1.056  &7.67 $\pm 0.22$ $\pm$ 0.34\\
	4.242	& 55.5 & 70  & $4.0 $ & 66.0  & 26.4 & 0.839& 1.056  &7.35 $\pm 0.96$ $\pm$ 0.32\\
	4.258	& 825.7& 960 & $18.8$ & 941.2 & 26.9 & 0.839& 1.052  &6.94 $\pm 0.23$ $\pm$ 0.31\\
	4.308	& 45.3 & 40  & $1.0 $ & 39.0  & 26.5 & 0.838& 1.054  &5.32 $\pm 0.87$ $\pm$ 0.23\\
	4.358	& 541.4& 538 & $19.5$ & 518.5 & 26.4 & 0.837& 1.051  &5.97 $\pm 0.27$ $\pm$ 0.26\\
	4.387	& 55.3 & 54  & $4.0 $ & 50.0  & 26.7 & 0.836& 1.051  &5.58 $\pm 0.85$ $\pm$ 0.25\\
    4.416   &1029.6& 949 & $20.8$ & 928.2 & 27.0 & 0.836& 1.053  &5.49 $\pm 0.18$ $\pm$ 0.24\\
	4.600	& 566.9& 395 & $16.4$ & 378.6 & 25.8 & 0.832& 1.054  &4.27 $\pm 0.23$ $\pm$ 0.19\\
	\hline \hline
    \end{tabular}
  \end{center}
\end{table*}

\section{\texorpdfstring{Detector and Monte-Carlo Simulation}{Detector
    and MC Simulation}}

The BESIII detector\,\cite{Ablikim2010345} at the BEPCII
collider\,\cite{BEPCII} is a large solid-angle magnetic spectrometer
with a geometrical acceptance of 93\% of 4$\pi$. It has four main
components: 1) A small-cell, helium-based (60\% He, 40\%
$\mbox{C}_3\mbox{H}_8$) multilayer drift chamber (MDC) with 43 layers
providing an average single-hit resolution of 135\,$\mu$m, a
charged-particle momentum resolution in a 1.0\,T magnetic field of
0.5\% at 1.0\,GeV/$c$ and a $dE/dx$ resolution better than 6\%; 2) A
time-of-flight system (TOF) constructed of 5 cm thick plastic
scintillator, with 176 detectors of 2.4\,m length in two layers in the
barrel and 96 fan-shaped detectors in the end-caps. The barrel
(end-cap) time resolution of 80\,ps (110\,ps) provides a 2$\sigma$
$K/\pi$ separation for momenta up to $\sim$ 1.0\,GeV$/c$; 3) An
electromagnetic calorimeter (EMC) consisting of 6240 CsI(Tl) crystals
in a cylindrical structure (barrel) and two end-caps. The energy and
the position resolutions for 1.0\,GeV photon are 2.5\% (5\%) and 6\,mm
(9\,mm) in the barrel (end-caps), respectively; 4) A muon system (MUC)
consisting of resistive plate chambers in nine barrel
and eight end-cap layers, which provides a 2\,cm position resolution.

To study the backgrounds and determine the detection efficiencies, a {\sc
geant4}-based\,\cite{GEANT4} Monte-Carlo (MC) simulation package is used,
which includes the geometric and material description of the \BESIII detector, the
detector response, and the digitization models, as well as the
detector running conditions and performance. Signal MC
samples of $e^+e^- \to K_{S}^{0}K^{+}\pi^{-}$ are generated with
phase space (PHSP) distributions with {\sc{evtgen}}\,\cite{EVTGEN_1,
EVTGEN_2},
which includes ISR effects~\cite{ISR_Formula}.  The
PHSP signal MC samples are reweighted according to the results from
the partial wave analysis (PWA) presented later in the paper.  For the
ISR calculation, the $e^+e^- \to K_{S}^{0}K^{+}\pi^{-}$ Born
cross-section results from BABAR\,\cite{BaBar} are taken as the
initial input, and the energy of the ISR photon is required to be less
than 0.1~GeV since the events with large energy ISR photons cannot
survive the event selection.  For the background study, an inclusive
MC sample with integrated luminosity equivalent to data is generated,
including open charm, low-mass vector charmonium states produced by
ISR, continuum light quark states, and other quantum electrodynamics
(QED) processes. The known decay modes of the charmonium states are
produced with {\sc{evtgen}}\,\cite{EVTGEN_1, EVTGEN_2} according to
the world average branching fraction (BF) values from the Particle
Data Group (PDG)\,\cite{PDG}, while the unknown decay modes are
generated with the {\sc lundcharm} generator\,\cite{LUNDCHARM}.

\section{Data analysis}
\label{PartII}
The signal candidates of the $e^+e^- \to K_{S}^{0}K^{+}\pi^{-}$
process are selected by requiring a $K_{S}^{0}$ candidate and a kaon
and pion pair with a net charge of zero.

The charged kaon and pion candidates, reconstructed using hits in the MDC,
are required to be within the polar angle range $|\cos\theta|<0.93$
and pass within a cylindrical region extending $\pm 10$~cm from the
average interaction point~(IP) of each run along the beam direction
and with a $1$~cm radius perpendicular to the beam direction.  The
time information from the TOF and the ionization measured in the
MDC~($dE/dx$) are combined to calculate particle identification~(PID)
confidence levels~(C.L.) for the $K$ and $\pi$ hypotheses, and the
particle type with the highest C.L. is assigned to each track.  An
identified kaon and an identified pion with opposite electric charge are
required.

The $K_{S}^{0}$ candidate is reconstructed with a pair of oppositely
charged tracks, which are assumed to be pions. Their distances of
closest approach to the IP must be within 25 cm and 20 cm along the
beam direction and in the transverse plane, respectively.  Then
primary and secondary vertex fits\,\cite{SecondVFit} are performed,
and the decay length of the secondary vertex is required to be greater
than twice its uncertainty. The invariant mass of $\pi^{+}\pi^{-}$,
$m_{\pi^{+}\pi^{-}}$, must satisfy
$|m_{\pi^{+}\pi^{-}}-M_{K_{S}^{0}}|< 0.020$~GeV/$c^{2}$, where
$M_{K_{S}^{0}}$ is the world average of the $K_{S}^{0}$
mass~\cite{PDG}.  To suppress the background from photon conversion,
the pions from the $K_{S}^{0}$ decay must satisfy $E/P$c$<0.8$, where $E$
and $P$ are the energy deposited in the EMC and the momentum measured
in the MDC, respectively.  If there are multiple $K_{S}^{0}$
candidates in an event, the one with the smallest $\chi^{2}$ of the
secondary vertex fit is taken.

To improve the momentum resolution and suppress background,
a four constraint (4C) kinematic fit is performed by imposing energy-momentum
conservation under the $e^+ e^- \to K_{S}^{0}K^{+}\pi^{-}$ hypothesis,
and its chi-square is required to be less than 40.

\begin{figure}[htbp]
  \includegraphics[width=0.35\textwidth]{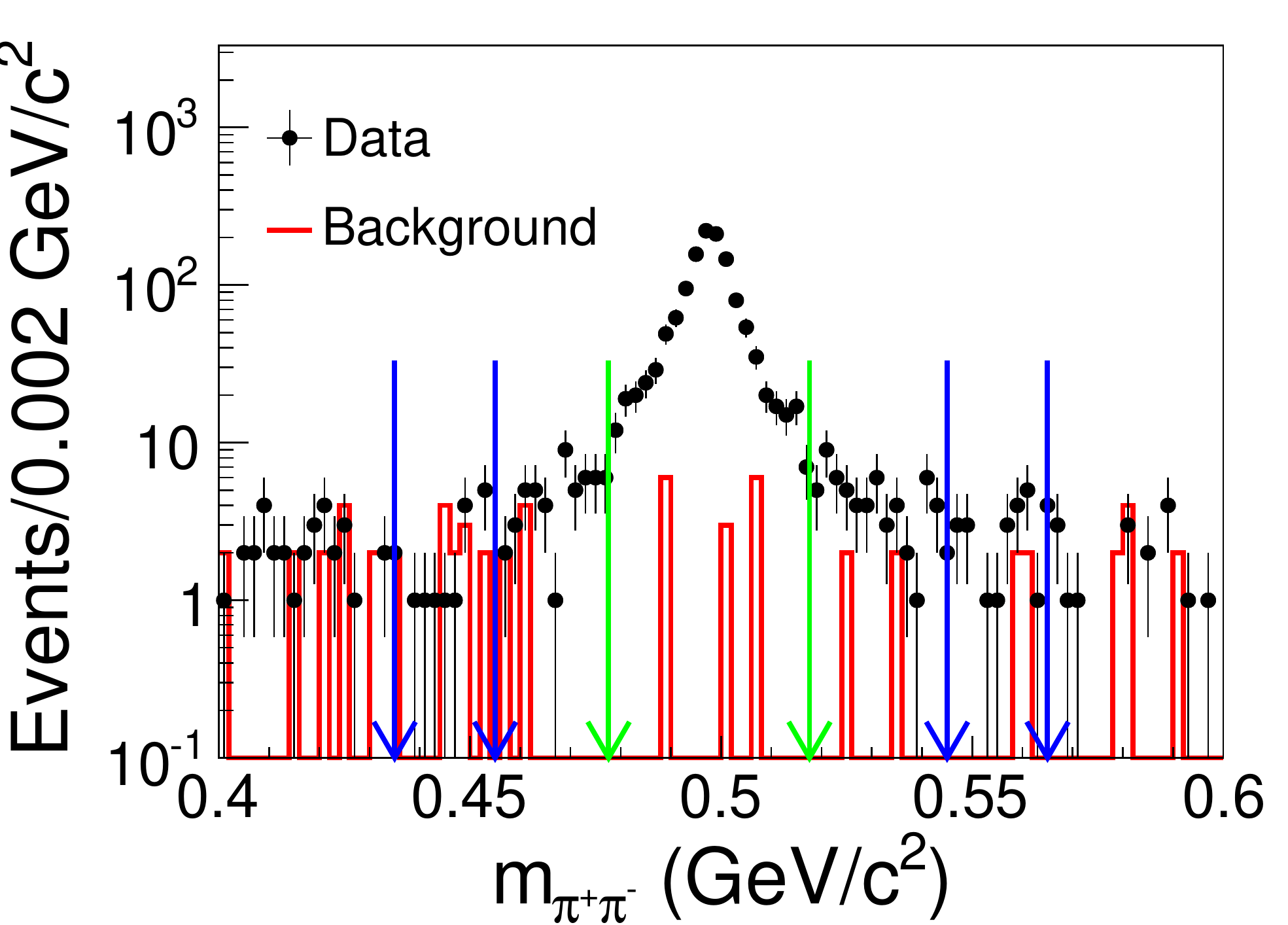}
  \caption{(Color online) The distribution of the $\pi^{+}\pi^{-}$
  invariant mass for the data at $\sqrt{s}=$ 4.226~GeV.
  The black dots with error bars are data, and the red
  histogram is background estimated from MC simulation. The blue
  arrows denote the sideband regions and green arrows shows
  the signal regions.}
\label{Ks_sideband}
\end{figure}

After all the event selection criteria are applied, the inclusive MC
sample shows that the surviving background is found to be mainly from
processes with 1) four charged tracks in the final state, $e.g.$,
$e^+e^- \to K^{+}K^{-}\pi^{+}\pi^{-}$, due to particle
misidentification between the kaon and pion and 2) a radiative photon,
$e.g.$, $e^+e^- \to \gamma e^+e^-$, which converts into an
electron-positron pair and the electron and positron are misidentified
as a pion and a kaon.
The signal yields, $N^{\rm{sig}}$, are obtained by counting
the events in the signal region $|m_{\pi^{+}\pi^{-}}-M_{K_{S}^{0}}|< 0.020$~GeV
and the number of remaining background events,
$N^{\rm{bkg}}$, is evaluated using the events in the sideband regions, which
are defined as $m_{\pi^{+}\pi^{-}}\in(0.435, 0.455) \cup (0.545, 0.565)$~GeV/$c^{2}$,
as shown in Fig.~\ref{Ks_sideband}. In the sideband region, there is still a small contribution
from signal events, which is estimated with signal MC simulation and subtracted
in the estimation of backgrounds.


\begin{figure}[!htbp]
    \subfigure{
        \includegraphics[width=0.3\textwidth]{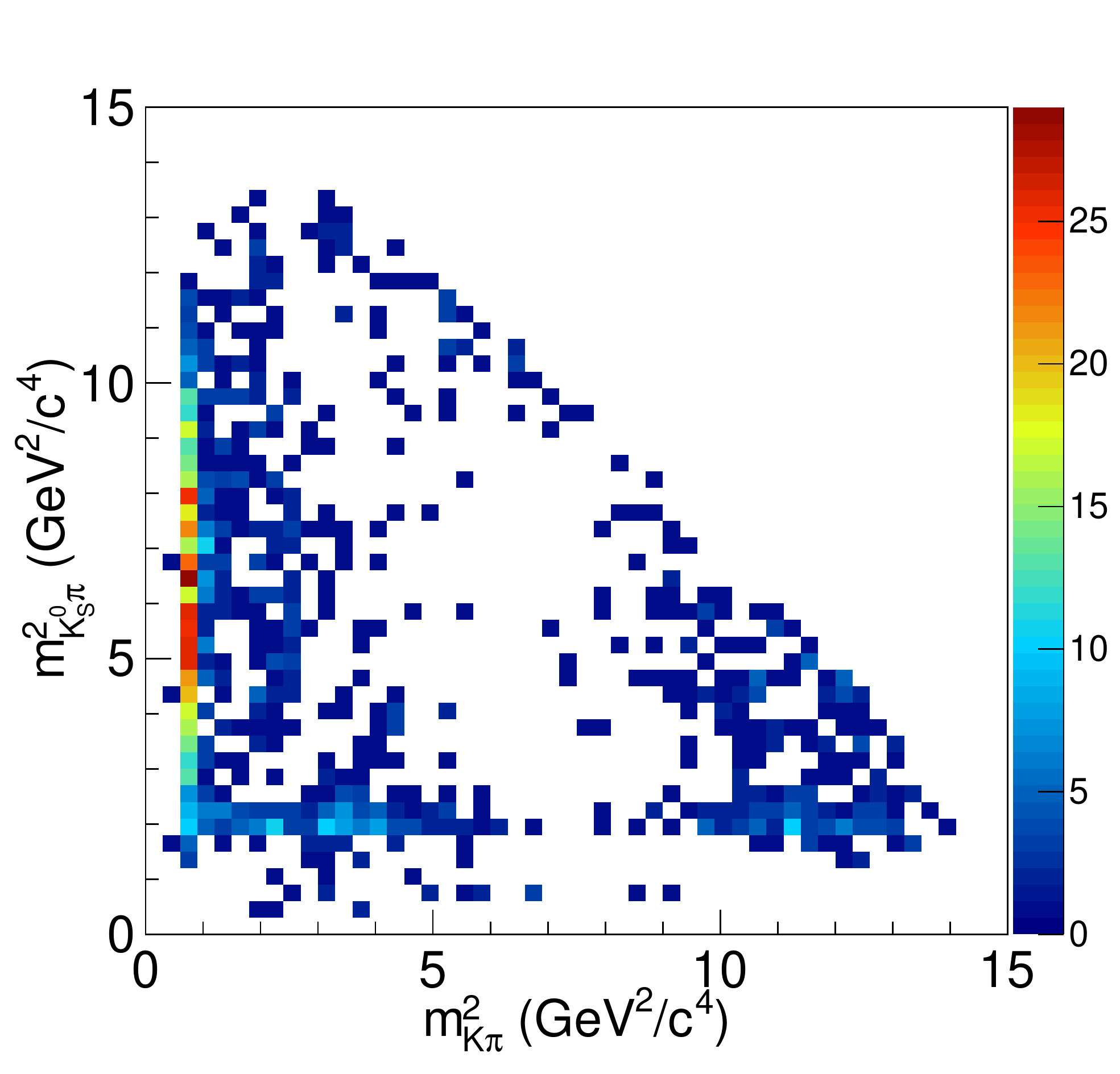}
        }
    \subfigure{
        \includegraphics[width=0.3\textwidth]{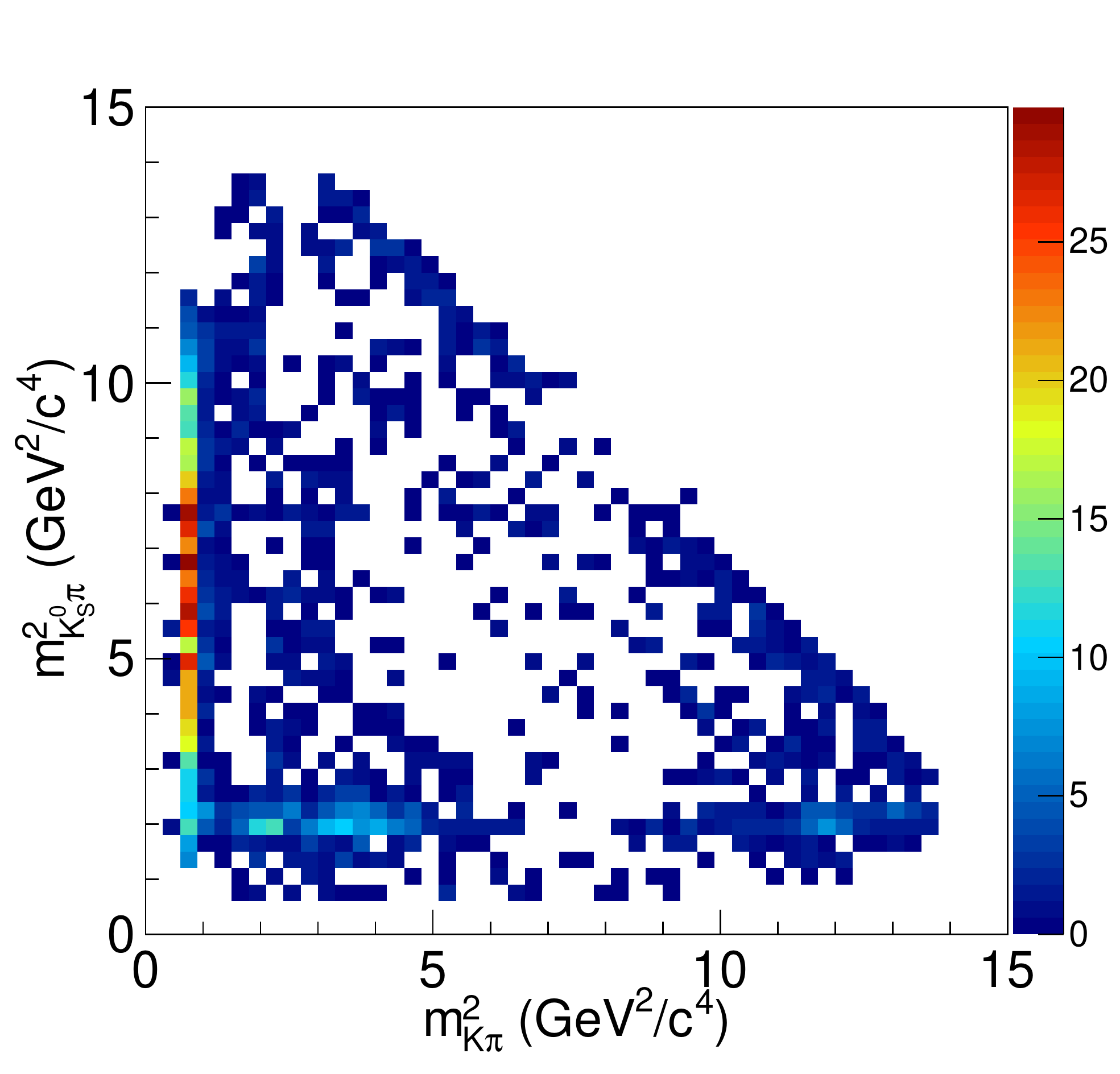}
        }
        \caption{The Dalitz plots of $e^{+}e^{-}\to
          K_{S}^{0}K^{+}\pi^{-}$ for the data at $\sqrt{s}=$
          4.226~GeV. The top plot is data and the bottom one is MC
          simulation generated with the amplitude analysis results.}
\label{KsKpi_DLT}
\end{figure}

\begin{figure*}[htbp]
    \subfigure{
        \includegraphics[width=0.32\textwidth]{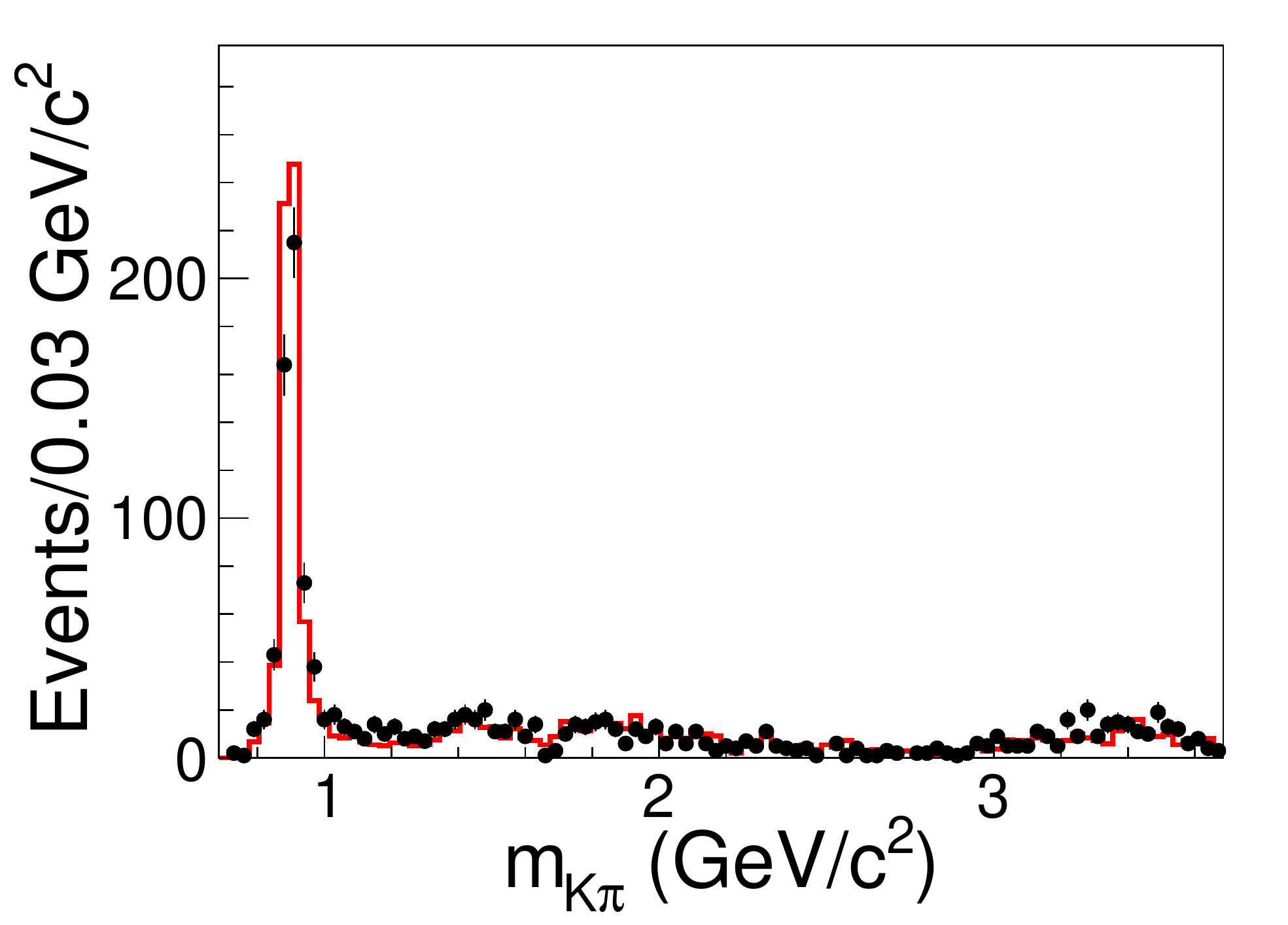}\put(-22,103){\bf (a)}
        }
    \subfigure{
        \includegraphics[width=0.32\textwidth]{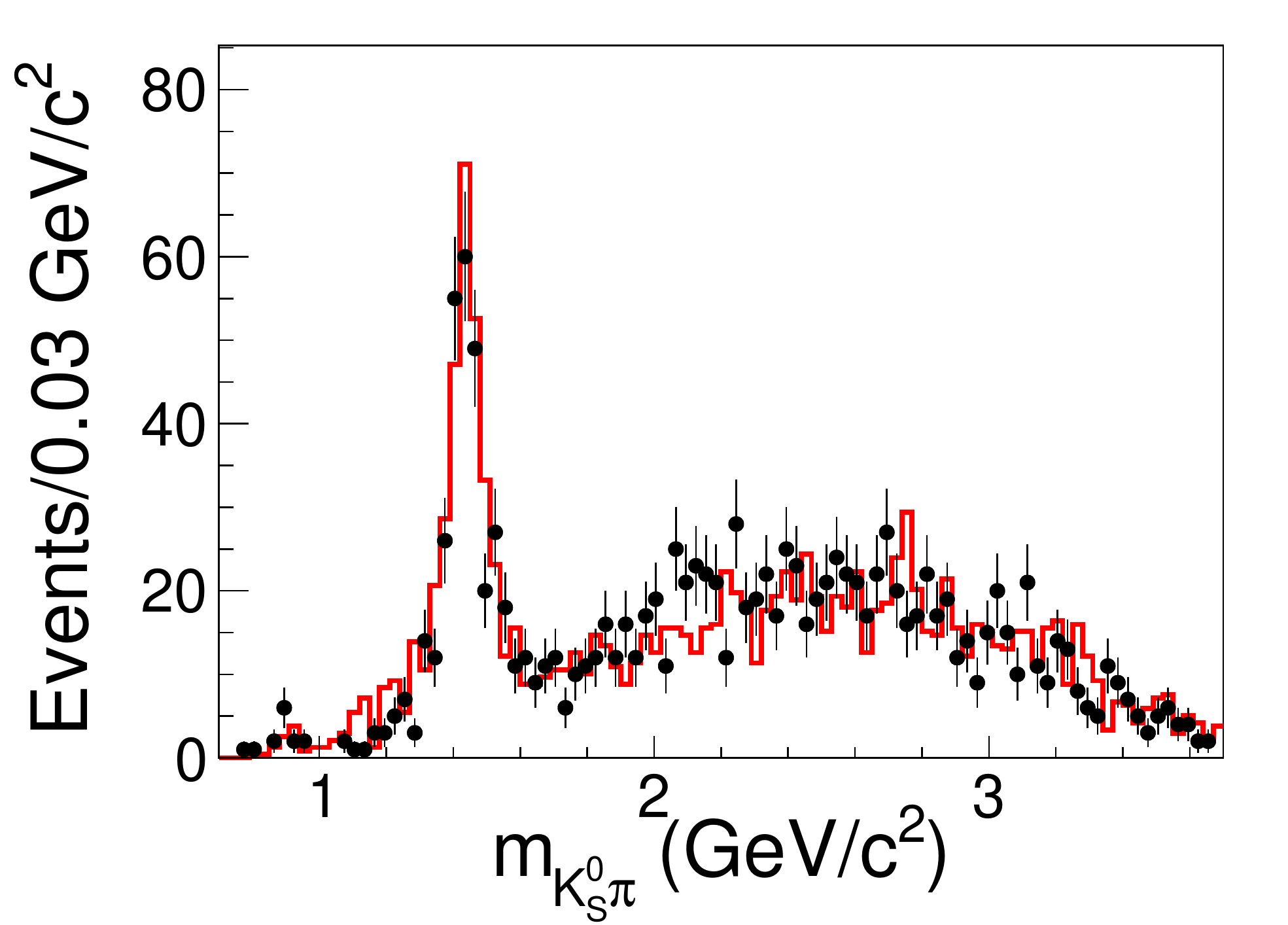}\put(-22,103){\bf (b)}
        }
    \subfigure{
        \includegraphics[width=0.32\textwidth]{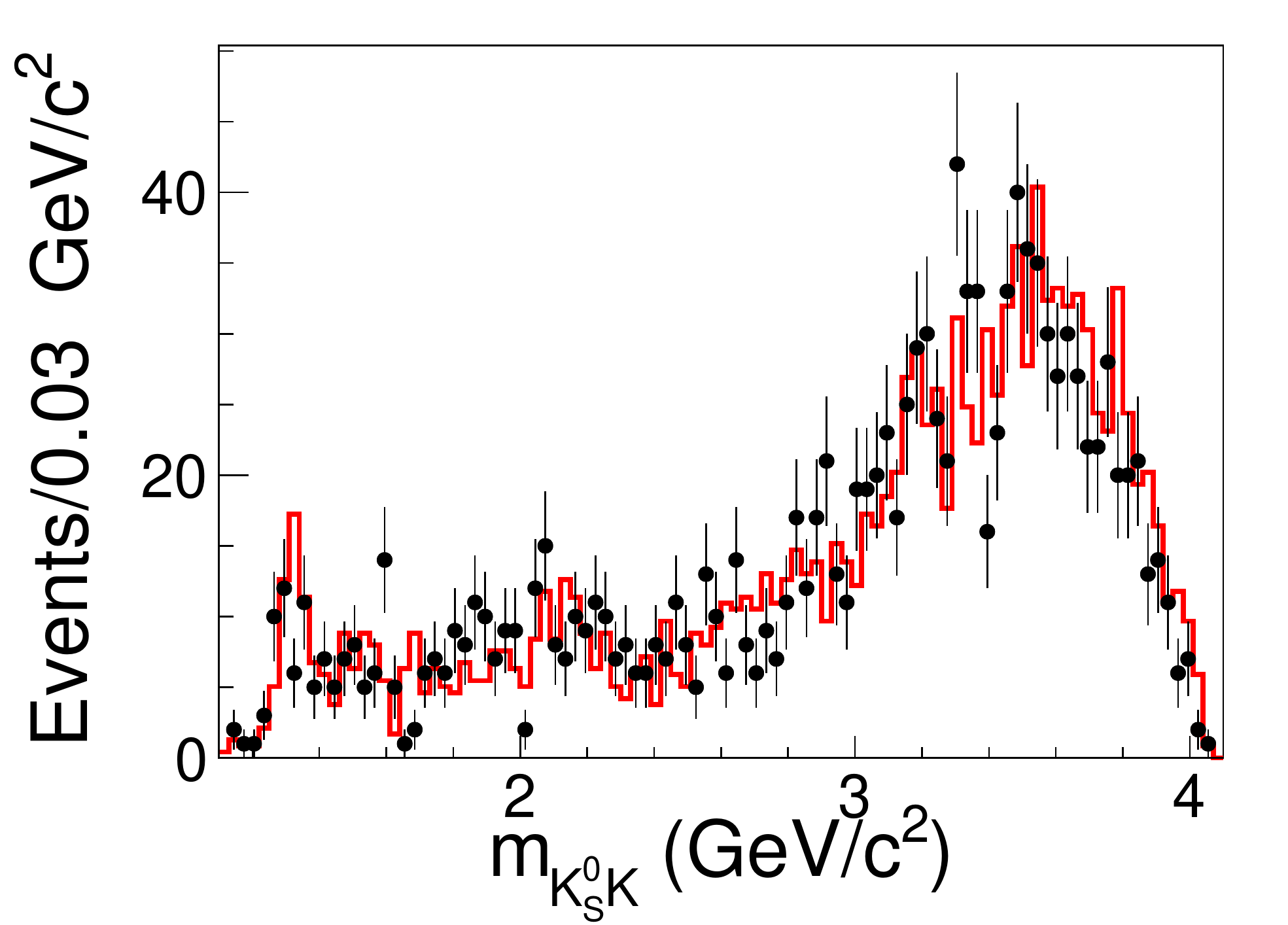}\put(-22,103){\bf (c)}
        }

    \subfigure{
        \includegraphics[width=0.32\textwidth]{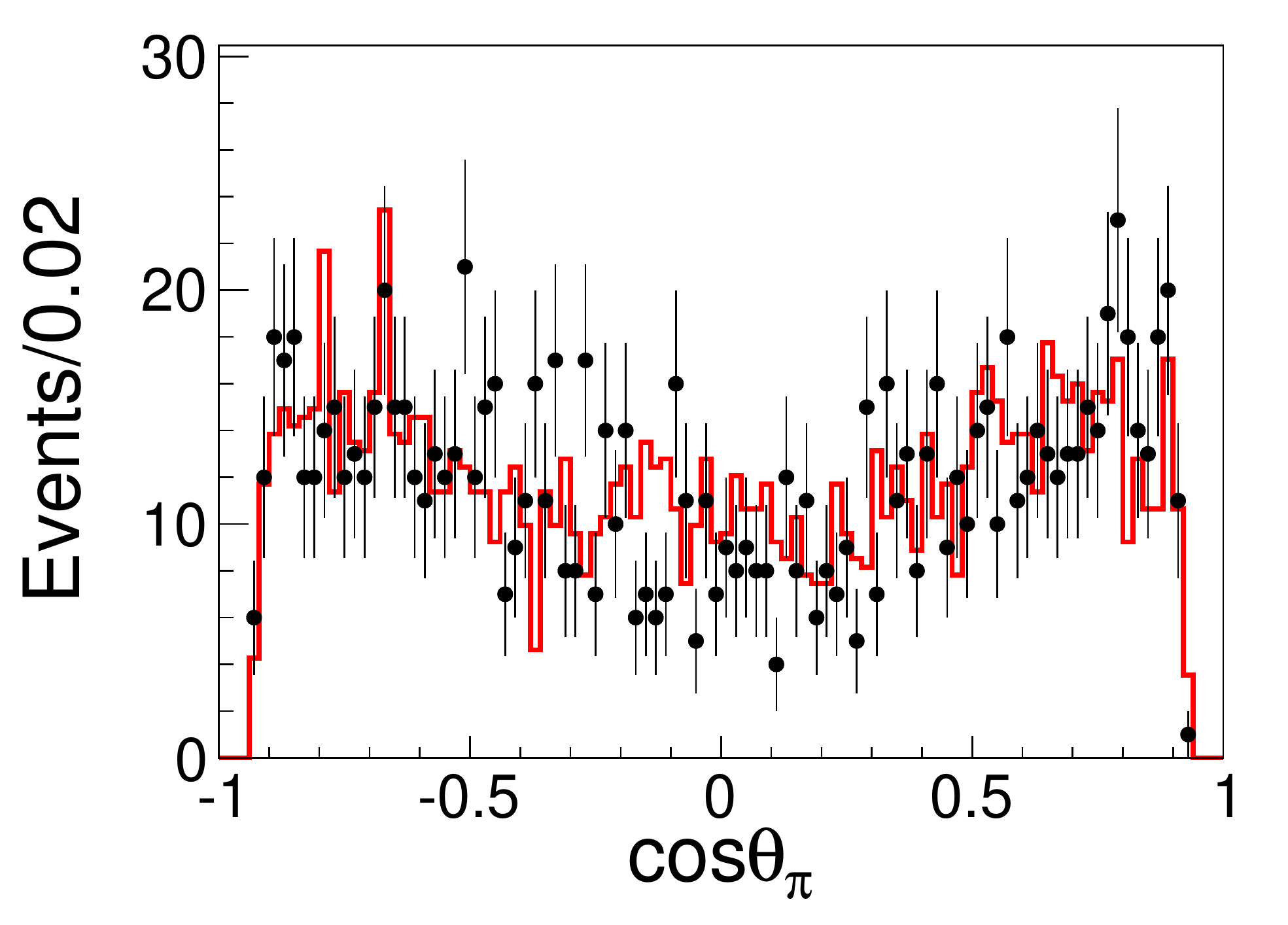}\put(-22,103){\bf (d)}
        }
    \subfigure{
        \includegraphics[width=0.32\textwidth]{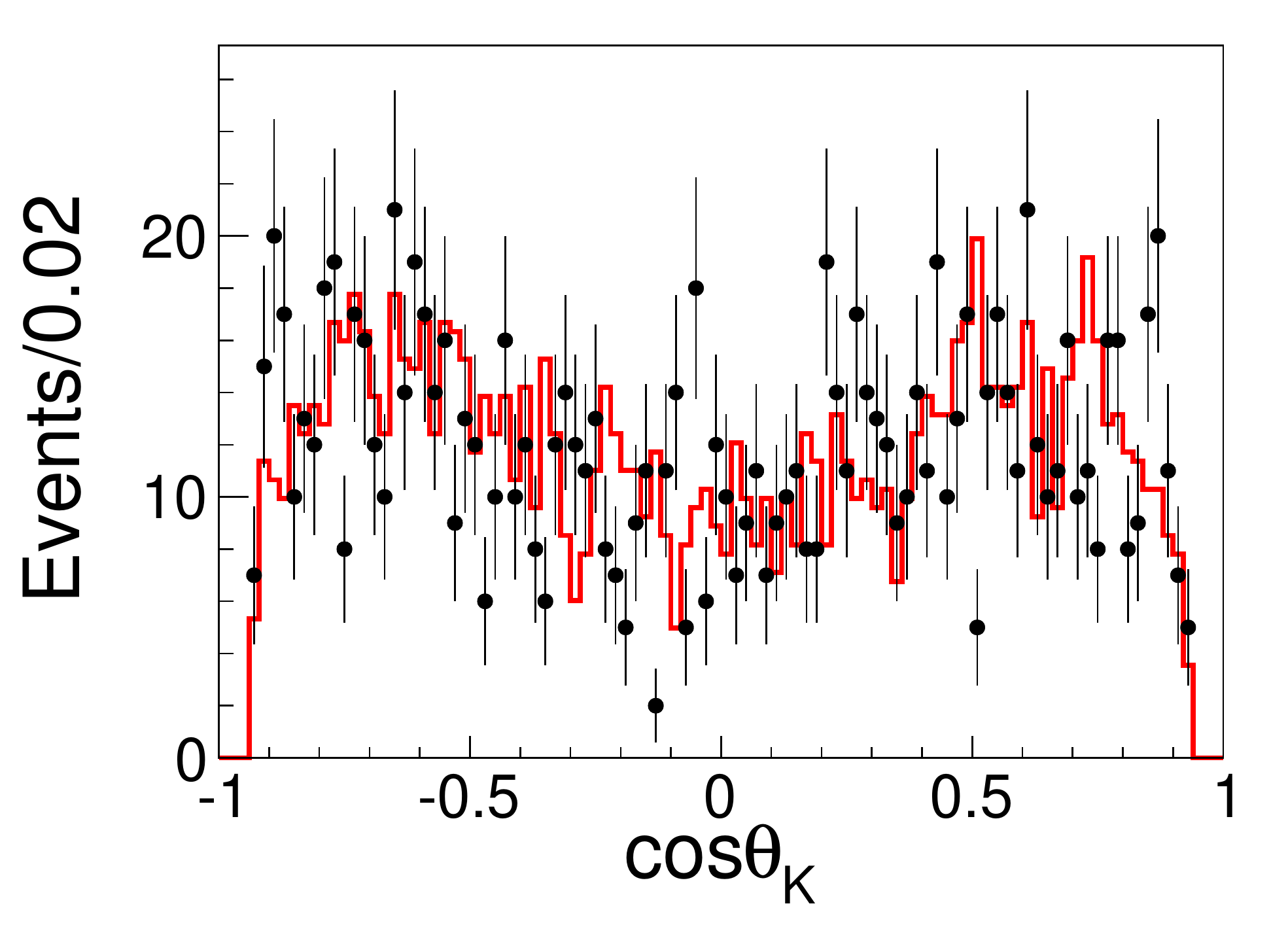}\put(-22,103){\bf (e)}
        }
    \subfigure{
        \includegraphics[width=0.32\textwidth]{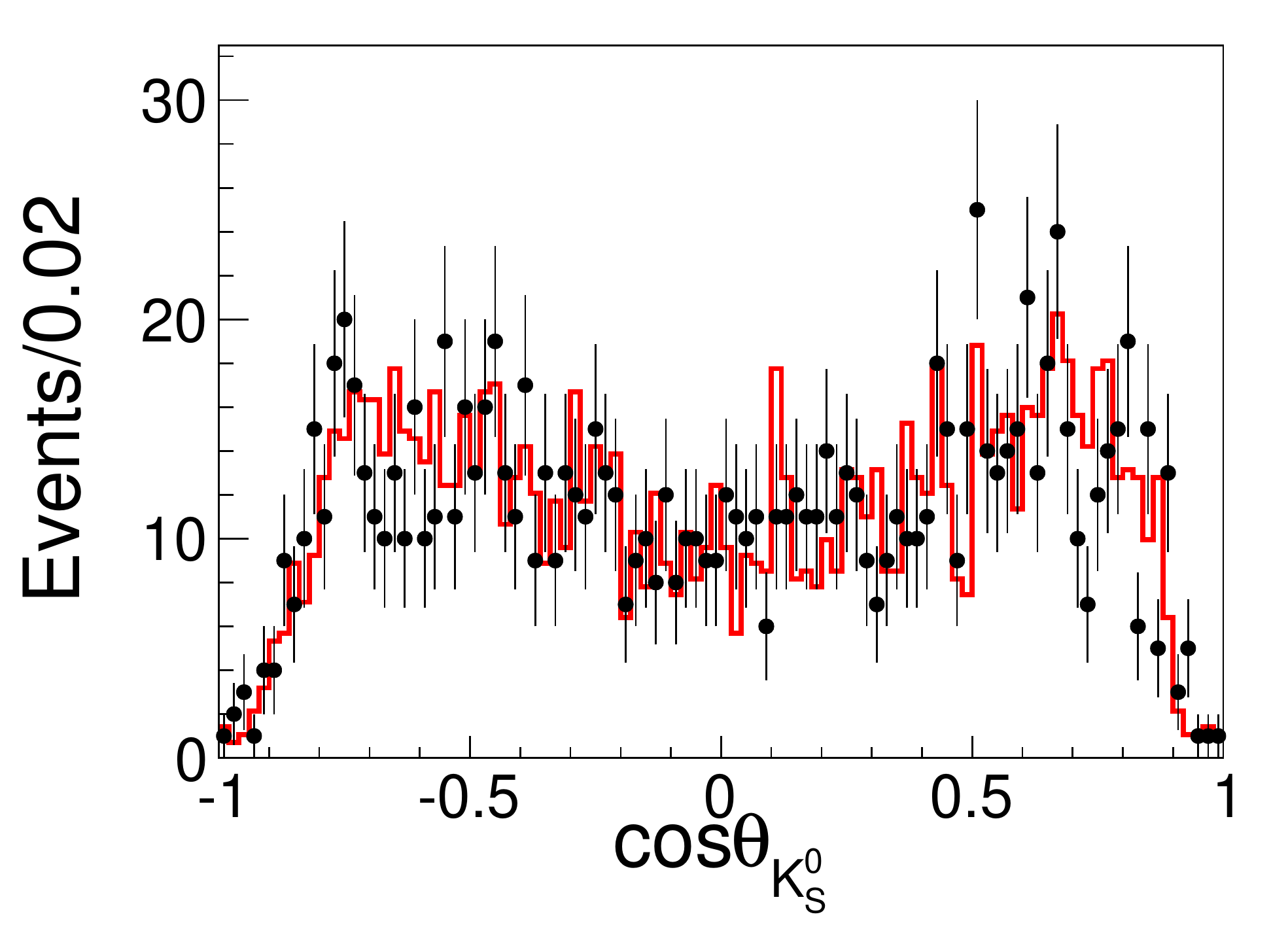}\put(-22,103){\bf (f)}
        }
        \caption{(Color online) Comparisons between data and MC
          simulation at $\sqrt{s} = $ 4.226~GeV.  The plots (a)-(c)
          are the invariant mass of $K\pi$, $K_{S}^{0}\pi$ and
          $K_{S}^{0}K$, and the plots (d)-(f) are the polar angle
          distributions of $\pi$, $K$ and $K_{S}^{0}$,
          respectively. Dots with error bars are data, and the red
          histograms are the MC projections from the amplitude
          analysis results.}
\label{KsKpi_PWA}
\end{figure*}

Figure~\ref{KsKpi_DLT}\,(top) shows the Dalitz plot of the selected events
at c.m. energy $\sqrt{s}=4.226$\,GeV.
Two vertical bands, corresponding to the neutral $K^{*}(892)$ and
$K^{*}_{2}(1430)$ decaying into $K^{\pm}\pi^{\mp}$, and a horizontal
band, corresponding to the charged $K^{*}_{2}(1430)$ decaying into
$K_{S}^{0}\pi^{\pm}$, are observed. There are also diagonal bands
corresponding to the intermediate states, $e.g.$ $a_{2}(1320)^{\pm}$
and excited $\rho^{\pm}$ with high mass, decaying into
$K_{S}^{0}K^{\pm}$.  In order to obtain the detection efficiencies,
PWAs are performed on the $K^{0}_{S}K\pi$
system at different c.m. energy points. The contributions of PHSP and
possible intermediate states in the $K_{S}^{0}\pi$, $K\pi$ and
$K_{S}^{0}K$ systems, including $K^{*}(892)$, $K_{2}^{*}(1430)$,
$K_{3}^{*}(1780)$, $a_{2}(1320)$, $\rho(1700)$ and $\rho(2150)$, are
taken into account. In the PWAs, these intermediate states are
described with relativistic Breit-Wigner~(BW) functions with their masses and widths fixed
to the world averages\,\cite{PDG}. The amplitudes for the
subsequent two body decays are constructed with the covariant helicity
method\,\cite{Helicity_1, Helicity_2}.  For a particle decaying into
a two-body final state, $i.e.$ $A(J,m) \to B(s,\lambda)C(\sigma,\nu)$,
its helicity amplitude $F_{\lambda, \nu}^{J}$\,\cite{Helicity_1, Helicity_2} is
\begin{equation}
  \begin{aligned}
  F_{\lambda,\nu}^{J} = \sum_{LS}\sqrt{\frac{2L+1}{2J+1}}g_{LS}\langle L\alpha S\delta|J\delta \rangle
                    \langle s\lambda\sigma - \nu|S\delta \rangle r^{L}\frac{B_{L}(r)}{B_{L}(r_{0})},
  \end{aligned}
  \label{Eq2}
\end{equation}
where $J$, $s$, and $\sigma$ are the spins of $A$, $B$, and $C$,
respectively; $m$, $\lambda$, and $\nu$ are their helicities,
respectively; $L$ and $S$ are the total orbital angular momentum and
spin of $AB$ system, respectively; $\alpha=0$; $\delta = \lambda -
\nu$; $g_{LS}$ is the coupling constant in the $L-S$ coupling scheme;
the angular brackets denote Clebsch-Gordan coefficients; $r$ is the
magnitude of the momentum difference between the two final state
particles in their mother's rest frame ( $r_{0}$ corresponds to the
momentum difference at the nominal mass of the resonance); and $B_{L}$
is the barrier factor\,\cite{Barrier}.  The magnitudes and relative
phases of complex coupling constants $g_{LS}$ are determined by an
unbinned maximum likelihood fit to data with
{\sc{minuit}}\,\cite{MINUIT}, and the effect of backgrounds is
subtracted from the likelihood as described in
Ref.\,\cite{KsKpi_BKG}. Figure~\ref{KsKpi_PWA} shows the fit results
for the invariant mass distributions of $K\pi$, $K_{S}^{0}\pi$, and
$K_{S}^{0}K$, as well as the polar angle distributions of $\pi$, $K$,
and $K_{S}^{0}$ at $\sqrt{s} = $ 4.226~GeV, where good agreement with
data is seen.  The situation of other data sets are similar.  Then the
detection efficiency $\epsilon$ is obtained by reweighting the signal
PHSP MC sample of $e^+e^- \to K_{S}^{0}K^{+}\pi^{-}$ with the fitted
PWA amplitude,
\begin{equation}
  \begin{aligned}
  \epsilon = \frac{\sum_{i=1}^{\rm{N_{MC}^{obs}}}|A_{i}|^{2}}{\sum_{i=1}^{\rm{N_{MC}^{gen}}}|A_{i}|^{2}},
  \end{aligned}
  \label{Eq3}
\end{equation}
where $\rm{N_{MC}^{gen}}$ and $\rm{N_{MC}^{obs}}$ are the numbers of generated
MC events and those passing the event selection, respectively, and $A_{i}$ is
the total amplitude of the $i$th event.

\begin{figure*}[htbp]
  \includegraphics[width=0.6\textwidth]{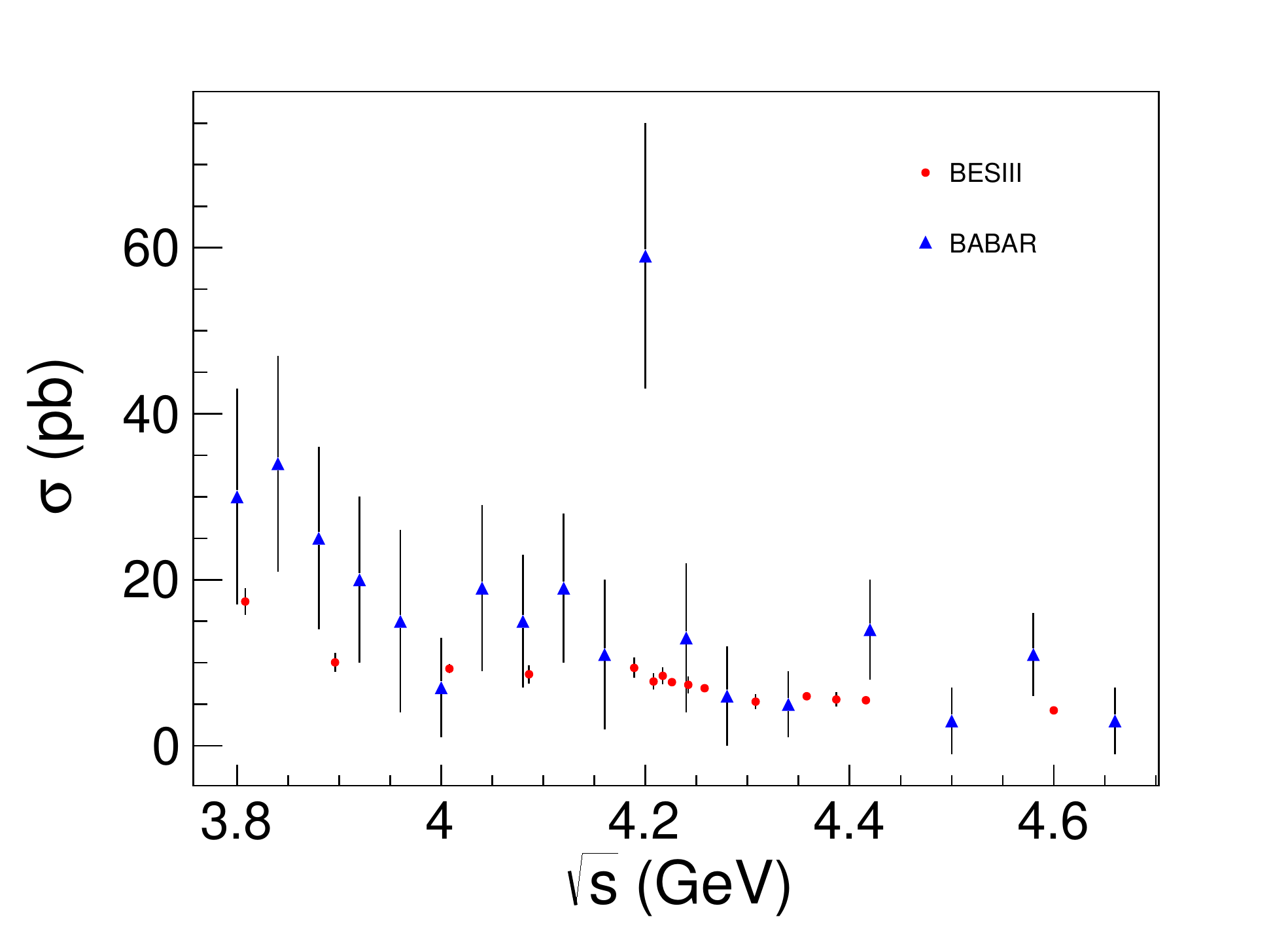}
  \caption{(Color online) The $e^+e^- \to K_{S}^{0}K^{+}\pi^{-}$ Born
    cross sections as a function of $\sqrt{s}$ (red dots) together with the
    previous results from the BABAR experiment\,\cite{BaBar} (blue triangles).
    Both statistical and systematic uncertainties are included.}
\label{Cross_section}
\end{figure*}

The Born cross sections are calculated with
\begin{equation}
  \begin{aligned}
  \sigma_{B}=\frac{N^{\rm{sig}}}{
    \mathcal{L}\times \mathcal{B}\times \epsilon\times
    (1+\delta^{\rm{ISR}})\times \frac{1}{|1-\Pi|^{2}}},
  \end{aligned}
  \label{Eq1}
\end{equation}
where $N^{\rm{sig}}$ is the signal yield with the subtraction of the
background contribution, $\mathcal{L}$ is
the integrated luminosity, $\mathcal{B}$ is the BF of the decay
$K_{S}^{0} \to \pi^{+}\pi^{-}$, $\epsilon$ is the detection
efficiency obtained by incorporating the PWA results as described
above, $(1+\delta^{\rm{ISR}})$ is the ISR correction factor, and
$\frac{1}{|1-\Pi|^{2}}$ is the vacuum polarization factor, which is taken from Ref.\,\cite{VacuumP}.
The ISR correction factor is obtained with
\begin{equation}
    1+\delta^{\rm{ISR}} = \frac{\sigma_{\rm{obs}}(s)}{\sigma_{B}(s)}=\frac{\int\sigma_{B}(s(1-x))\,F(x,s)\,dx}{\sigma_{B}(s)},
\end{equation}
where $\sigma_{\rm{obs}}$ is the observed cross section, $s$ is the square of c.m. energy, $x$ is the fraction
of the beam energy taken by the radiative photon, and $F(x,s)$ is the radiator function~\cite{ISR_Formula}.
To get the correct ISR photon energy distribution, the
cross section of $e^+e^- \to K_{S}^{0}K^{\pm}\pi^{\mp}$ measured by BABAR\,\cite{BaBar}
is taken as the input to get the initial ISR correction factor and cross section, the latter is added to re-calculate
the ISR correction factor. We repeat this process till both the ISR correction factors and cross section converge.
The measured Born cross sections for the
individual c.m. energy points are summarized in
Table~\ref{KsKpi_result}, as well as other quantities used to
calculate the Born cross section. A comparison of the Born cross
sections between our measurement and BABAR's results in the
c.m. energy region $\sqrt{s}=$ 3.800 $\sim$ 4.660~GeV is shown in
Fig.~\ref{Cross_section}. The measured cross sections
agree with but are of much higher precision than those obtained
by BABAR\,\cite{BaBar}.

\begin{figure}[htbp]

    \subfigure{
        \includegraphics[width=0.4\textwidth]{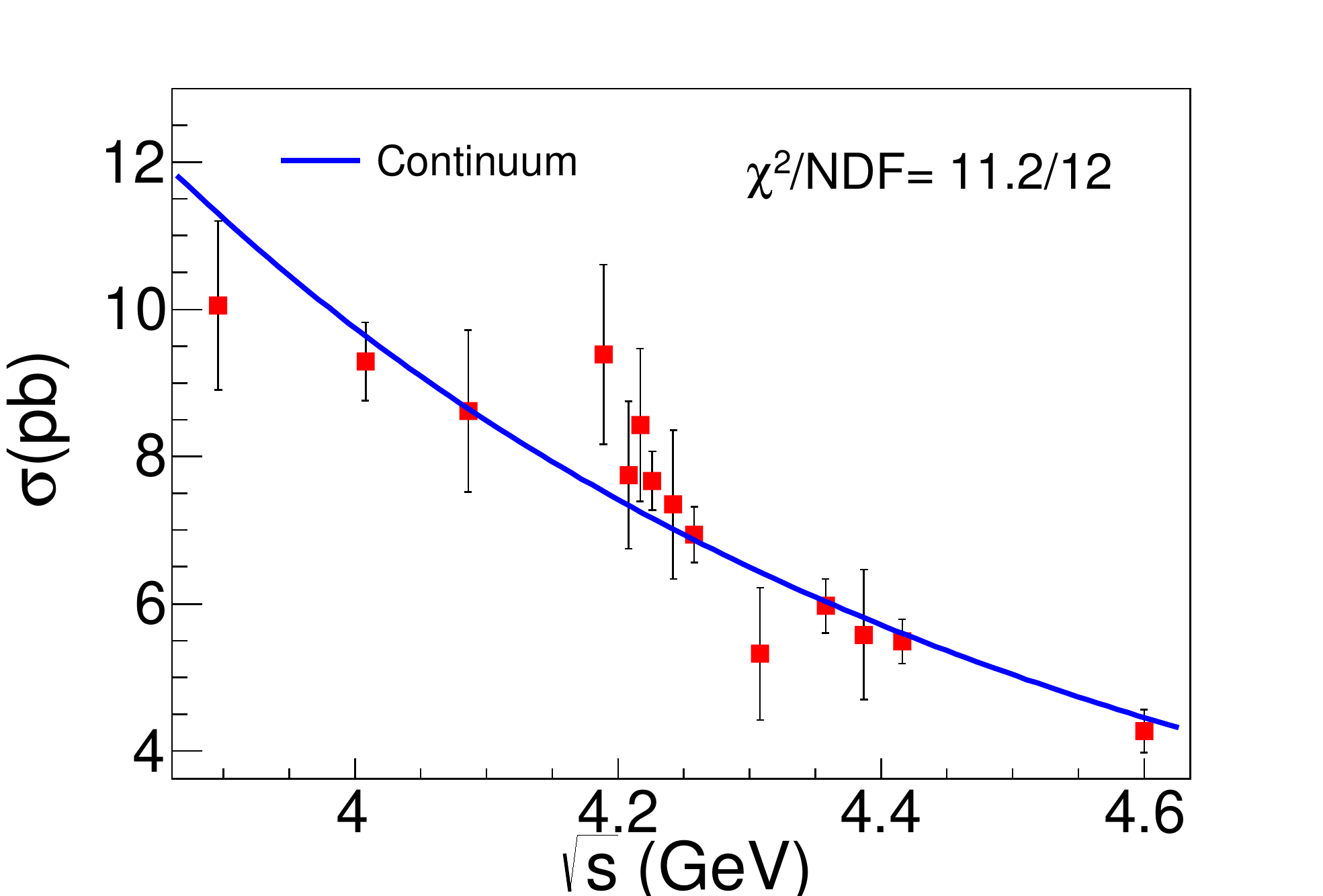}\put(-90,90){\bf (a)}
        }

    \subfigure{
        \includegraphics[width=0.4\textwidth]{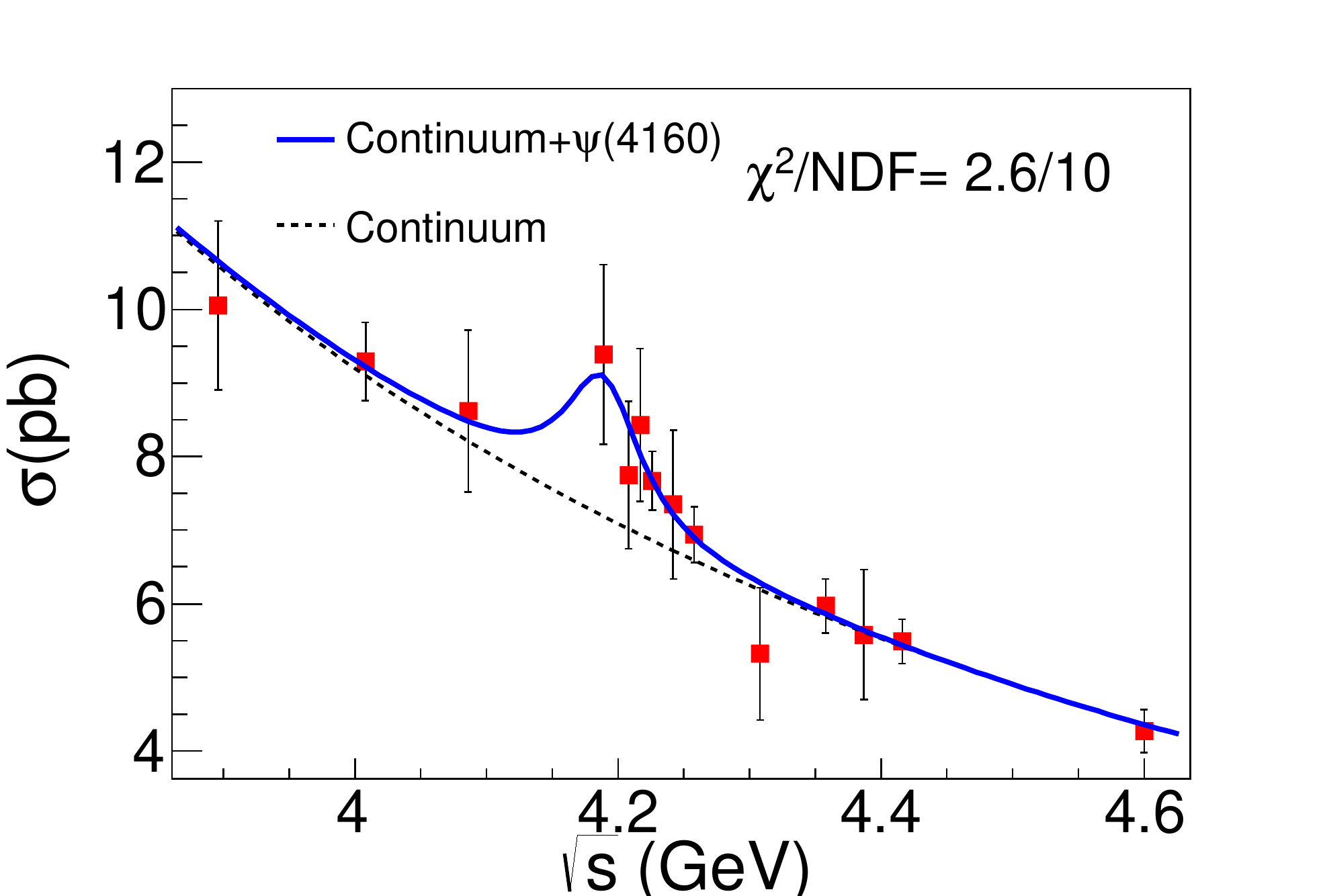}\put(-90,90){\bf (b)}
        }

    \subfigure{
        \includegraphics[width=0.4\textwidth]{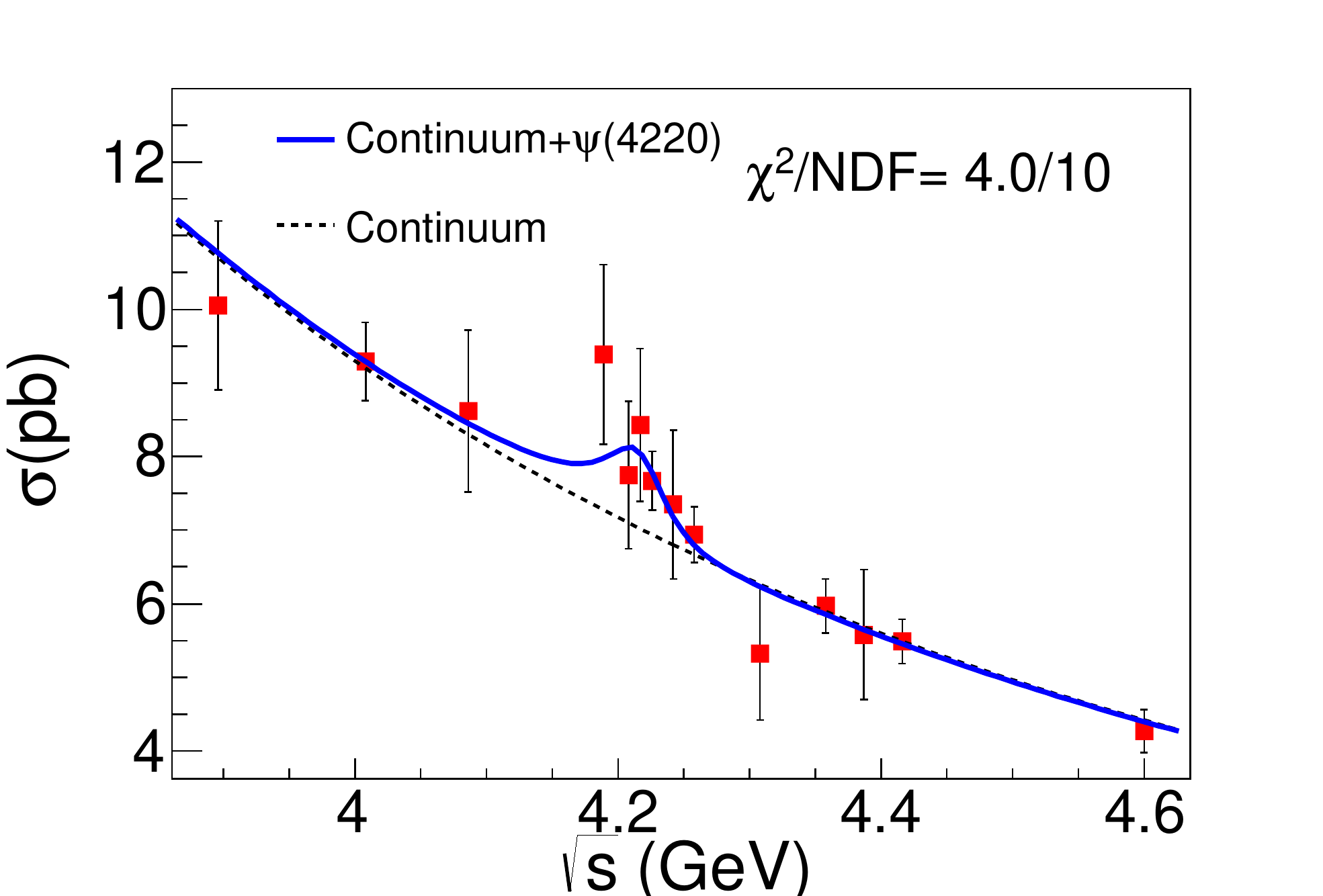}\put(-90,90){\bf (c)}
        }
        \caption{(Color online) Fit to the $\sigma_{B}(e^+e^- \to
          K_{S}^{0}K^{+}\pi^{-})$ Born cross section.  The data (red
          squares) include both statistical and systematic
          uncertainties, the  solid curves are the projections from the
          best fit, and the dashed curves show the fitted continuum
          components.  The top plot is the result with continuum
          process only, the middle one is with continuum and
          $\psi(4160)$, and the bottom one is with continuum and
          $Y(4220)$.}
\label{Fig_CorrFit}
\end{figure}

The $e^{+}e^{-} \to K_{S}^{0}K^{+}\pi^{-}$ Born cross sections of this
work are fitted with a $1/s^{n}$ function. BABAR's\,\cite{BaBar}
results have large uncertainties above 3.8~GeV, so they are not
included.  In addition, the data point at around 3.8~GeV is not used
in the fit, since an attempt to fit the cross section around this
energy should consider the contribution from $\psi(3770)$. There is
only one data point close to the $\psi(3770)$ peak, which is
insufficient to constrain the parameters associated with $\psi(3770)$.
The correlations among different data points are considered in the
fit, with the chi-square function constructed as Eq.~\ref{Fit_function}, which is
minimized by {\sc{minuit}}\,\cite{MINUIT},
\begin{equation}
    \chi^{2}= \sum_{i}\frac{(\sigma_{B_{i}}-h\cdot\sigma_{B_{i}}^{\rm{fit}})^2}
    {\delta_{i}^{2}} + \frac{(h-1)^2}{\delta_{c}^2}.
\label{Fit_function}
\end{equation}
Here, $\sigma_{B_{i}}$ and $\sigma_{B_{i}}^{\rm{fit}}$ are the measured
and fitted Born cross sections of the $i$th energy point, respectively;
$\delta_{i}$ is the independent part of the total uncertainty,
which includes the statistical uncertainty and the uncorrelated part of
the systematic uncertainty~(the details are in Sec.~\ref{Sys uncertainty});
$\delta_{c}$ is the correlated part of the systematic uncertainty, which
will be described in detail in the next section;
and $h$ is a free parameter introduced to take into account the correlations.
Figure~\ref{Fig_CorrFit}(a) shows the fit result with a goodness-of-the-fit of
$\chi^{2}/{\rm NDF}=11.2/12$, where the solid curve shows the continuum process.
A better fit is obtained by using the coherent sum of the continuum
and the $\psi(4160)$ or $Y(4220)$ amplitude (the two closest states around
the excess of the cross section). The fit function used is
\begin{equation}
    \sigma = \left |\sqrt{\frac{f_{\rm{con}}}{s^{n}}} + e^{i\phi}\frac{\sqrt{12\pi
    \Gamma_{e^{+}e^{-}}B_{K_{S}^{0}K\pi}\Gamma}}{s-M^{2}+iM\Gamma }\right|^{2},
\end{equation}
where $f_{\rm{con}}$ and $n$ are the fit parameters for the continuum
process, $\phi$ is the relative phase between the continuum and
resonant amplitudes, $\Gamma$ and $\Gamma_{e^{+}e^{-}}$ are the width
and partial width to $e^{+}e^{-}$, respectively, $B_{K_{S}^{0}K\pi}$
is the BF of the resonance decays into $K_{S}^{0}K^{+}\pi^{-}$, and
$M$ is the mass of the resonance. The masses and total widths of
$\psi(4160)$ and $Y(4220)$ are fixed to Refs.~\cite{PDG, Y(4220)}.
Two solutions with the same minimum value of $\chi^{2}$ are found with
different interference between the two amplitudes.  The fit results
are shown in Figs.~\ref{Fig_CorrFit} (b) and (c) (the lineshapes of the two solutions
are identical) and summarized in Table II.  The corresponding
significance for $\psi(4160)$ is 2.5$\sigma$ and for $Y(4220)$
2.2$\sigma$.

\begin{table*}[htbp]
\caption{Results of the fits to the Born cross section $\sigma_{B}$.
     Shown in the table are the product of the $e^{+}e^{-}$ partial width and
     the BF to the $K_{S}^{0}K^{+}\pi^{-}$ final state $\Gamma_{e^{+}e^{-}}\times B_{K_{S}^{0}K^{+}\pi^{-}}$,
     the relative phase between the different amplitudes $\phi$,
     and the corresponding significance of $\psi(4160)$ and $Y(4220)$.
     The uncertainties of the parameters are from the fits.}
\label{Tab_CorrFit}

\begin{tabular}{c|c|c|c|c}
    \hline \hline
        \multirow{2}{*}{}&\multicolumn{2}{c|}{$\psi(4160)$} & \multicolumn{2}{c}{$Y(4220)$}\\
        \cline{2-5}
        & Solution I & Solution II & Solution I & Solution II\\
        \hline
        $\Gamma_{ee}\times B_{K_{S}^{0}K^{+}\pi^{-}}$~(eV) &2.71$\pm$0.13 & 0.0095$\pm$0.0088 &2.04$\pm$0.19& 0.0027$\pm$0.0023\\
        $\phi$ (rad)& -1.60$\pm$0.03 & 1.67$\pm$0.44 & -1.60$\pm$0.02 & 2.00$\pm$0.53\\
        \hline
        Significance & \multicolumn{2}{c|}{$2.5\sigma$} &\multicolumn{2}{c}{$2.2\sigma$}\\
        \hline \hline
\end{tabular}
\end{table*}

\section{Systematic Uncertainties}
\label{Sys uncertainty}

Various sources of systematic uncertainties are investigated for the
cross section measurements of $e^+e^- \to K_{S}^{0}K^{+}\pi^{-}$,
and all of them are summarized in Table~\ref{total_sys}.

\begin{table}[htbp]
	\caption{Systematic uncertainties of the measurements of $\sigma(e^+e^- \to K_{S}^{0}K^{+}\pi^{-})$.}
	\label{total_sys}
	\begin{center}
	\begin{tabular}{lc}
	\hline	\hline
    Source 	& Relative uncertainty (\%)\\
	\hline
    Tracking  & 2.0 \\
    PID  &2.0 \\
    $K_{S}^{0}$ reconstruction & 1.2\\
    Kinematic fit	& 0.5 \\
    Signal model    & 2.0 \\
    Signal yield   & 1.8 \\
    ISR factor          & 1.0 \\
    Integrated luminosity	& 1.0 \\
    BF  & 0.1 \\
	\hline
    Total				& 4.4 \\
	\hline	\hline
	\end{tabular}
	\end{center}
\end{table}

The systematic uncertainties associated with tracking and PID have
been studied using control samples of $J/\psi \to
\pi^{+}\pi^{-}p\bar{p}$ and $J/\psi \to K_{S}^{0}K^{\pm} \pi^{\mp}$
with $K_{S}^{0} \to \pi^{+}\pi^{-}$\,\cite{Tracking_PIDeff}, and
the kaon and pion tracking and PID efficiencies for data agree
with those of MC simulation within 1\%, so the total tracking
and PID uncertainties are both determined to be 2\%~(1.0\% per track).

The uncertainty associated with $K_{S}^{0}$ reconstruction is studied
with the processes $J/\psi \to K^{*\pm}K^{\mp}$ and $J/\psi \to \phi
K_{S}^{0}K^{\pm}\pi^{\mp}$\,\cite{Ks_eff}.  The difference of the
reconstruction efficiency between data and MC simulation is found to
be 1.2\%, which is taken as the systematic uncertainty.

The systematic uncertainty due to the kinematic fit is estimated by
correcting the track helix parameters of charged tracks and the
corresponding covariance matrix for the signal MC sample to improve
the agreement between data and MC simulation. The detailed method can be
found in Ref.\,\cite{refsmear}.  The resulting change of the detection
efficiency with respect to the one obtained without the
corrections is taken as the systematic uncertainty.

In the measurement of cross section for $e^+e^- \to
K_{S}^{0}K^{+}\pi^{-}$, the detection efficiency is estimated with the
weighted PHSP MC samples, where the weights are obtained according to
the PWA results. To estimate the corresponding systematic uncertainty
associated with the signal MC model, we repeat the PWA by 1) changing the
resonance parameters of the intermediate states by one standard
deviation~\cite{PDG} and by 2) excluding the intermediate state with the
least significance in the fit. The alternative PWA results are used to
recalculate the detection efficiency, and the resulting differences
are taken as the systematic uncertainties.  Assuming the two contributions
are uncorrelated, the overall uncertainty associated
with the signal MC model is the sum of the above individual values
in quadrature.  To minimize the effect of the limited statistics of data,
the uncertainty for the data sample at
$\sqrt{s}=4.226$~GeV, which has the largest integrated luminosity of
all the samples, is used, and the value, 2.0\%, is assigned to all
c.m. energy points.

For the systematic uncertainties associated with the signal yield
determinations, we repeat the analysis by changing the mass interval of
$M_{\pi^{+}\pi^{-}}$ from 0.03 to 0.04\,GeV/$c^{2}$, and by changing
the $K_{S}^{0}$ sideband regions to $m_{\pi^{+}\pi^{-}} \in (0.43,
0.45) \cup (0.55, 0.57)$~GeV/$c^{2}$.  The largest change of the
signal yields with respect to the nominal value among all c.m. energy
points, 1.8\%, is conservatively taken as the systematic uncertainty.

The uncertainty associated with the vacuum polarization
factor\,\cite{VacuumP} is negligible compared with the other
uncertainties. For the ISR correction factors, the iteration procedure
is carried out until the measured Born cross section converges. The
convergence criterion, 1.0\%, is taken as the systematic uncertainty.

The integrated luminosities at each c.m. energy point are measured
using large angle Bhabha scattering events with an uncertainty of
1.0\%\,\cite{Lumi}. The uncertainty on the BF of the decay $K_{S}^{0}
\to \pi^{+}\pi^{-}$ is from the PDG\,\cite{PDG}.

Assuming all sources of systematic uncertainties are uncorrelated, the
total systematic uncertainty is obtained by adding the individual
values in quadrature and are summarized in Table~\ref{total_sys}.

\section{Summary}
The $e^+e^- \to K_{S}^{0}K^{\pm}\pi^{\mp}$ Born cross sections have
been measured by BESIII at the c.m. energy region from $3.8$ to
$4.6$~GeV, and the results are shown in Fig.~\ref{Cross_section} and
summarized in Table~\ref{KsKpi_result}. The cross sections agree with
BABAR's results\,\cite{BaBar}, but with significantly improved
precision.  The line shape of the Born cross sections is consistent
with only the continuum process, however a better fit is obtained by
adding an additional resonance. The fit to the Born cross sections
from this work, with $\psi(4160)$~($Y(4220$)) added, is performed.
Only evidence for the $\psi(4160)$~($Y(4220)$) is observed with the
corresponding significance $2.5\sigma$~($2.2\sigma$).  Further study
of this channel with more energy points and larger statistics will be
essential for a deeper understanding of the line shape and
contributions from charmonium and charmonium-like states.

\begin{acknowledgments}
The BESIII collaboration thanks the staff of BEPCII and the IHEP computing center for their strong support. This work is supported in part by National Key Basic Research Program of China under Contract No. 2015CB856700; National Natural Science Foundation of China (NSFC) under Contracts Nos. 11235011, 11335008, 11425524, 11625523, 11635010; the Chinese Academy of Sciences (CAS) Large-Scale Scientific Facility Program; the CAS Center for Excellence in Particle Physics (CCEPP); Joint Large-Scale Scientific Facility Funds of the NSFC and CAS under Contracts Nos. U1332201, U1532257, U1532258; CAS under Contracts Nos. KJCX2-YW-N29, KJCX2-YW-N45; CAS Key Research Program of Frontier Sciences under Contracts Nos. QYZDJ-SSW-SLH003, QYZDJ-SSW-SLH040; 100 Talents Program of CAS; National 1000 Talents Program of China; INPAC and Shanghai Key Laboratory for Particle Physics and Cosmology; German Research Foundation DFG under Contracts Nos. Collaborative Research Center CRC 1044, FOR 2359; Istituto Nazionale di Fisica Nucleare, Italy; Koninklijke Nederlandse Akademie van Wetenschappen (KNAW) under Contract No. 530-4CDP03; Ministry of Development of Turkey under Contract No. DPT2006K-120470; National Science and Technology fund; The Swedish Research Council; U. S. Department of Energy under Contracts Nos. DE-FG02-05ER41374, DE-SC-0010118, DE-SC-0010504, DE-SC-0012069; University of Groningen (RuG) and the Helmholtzzentrum fuer Schwerionenforschung GmbH (GSI), Darmstadt; WCU Program of National Research Foundation of Korea under Contract No. R32-2008-000-10155-0.
\end{acknowledgments}


\end{document}